\documentclass[dvips,12pt]{emulateapj-rtx4}
\usepackage[english]{babel}
\usepackage{amsmath}
\usepackage{amssymb}
\usepackage{url}
\usepackage[dvips]{color}

\usepackage[dvipdfm,hypertexnames=false]{hyperref} 

\newcommand{\be}{\begin{equation}}
\newcommand{\ee}{\end{equation}}
\newcommand{\beq}{\begin{eqnarray}}
\newcommand{\eeq}{\end{eqnarray}}

\begin{document}

\title{Spin equilibrium with or without gravitational wave emission:\\
the case of XTE J1814-338 and SAX J1808.4-3658}
\author{B. Haskell, A. Patruno} 
\affiliation{Astronomical Institute ``Anton Pannekoek'', University of
  Amsterdam, Postbus 94249, 1090 GE Amsterdam, the Netherlands}

\begin{abstract}
\noindent
In this letter we present a new analysis of the torques acting on the
accreting millisecond X-ray pulsars SAX J1808.4-365 and XTE
J1814-338, and show how our results can be used to constrain
theoretical models of the spin evolution. In particular we find upper
limits on any spin-up/down phase of XTE J1814-338 of
$|\dot{\nu}|\lesssim\,1.5\times\,10^{-14}\rm\,Hz\,s^{-1}$ at 95\%
confidence level. We examine the possibility that a gravitational wave
torque may be acting in these systems and suggest that a more likely
scenario is that both systems are close to spin equilibrium,
as set by the disc/magnetosphere interaction.

\end{abstract}
\maketitle

\section{Introduction}

Accreting millisecond X-ray pulsars (AMXPs) are one of the best
laboratories to test our understanding of strong gravity and of matter
at extreme densities. In these systems a neutron star (NS) in a binary
accretes gas from a less evolved low mass donor companion. During the
accretion process angular momentum is transfered to the neutron star, which can be spun up to millisecond periods.

This picture is not, however, without its complications. As more AMXPs
and millisecond radio pulsars are discovered there appears to be strong evidence for a
cutoff in the distribution of the spin rates at approximately 730 Hz,
which is well below the Keplerian breakup frequency for these objects
\citep{Chak1,Ale1}. It thus appears that the torques acting on these
systems are stopping them from spinning up to the breakup limit.  A
detailed analysis of the torques acting on an AMXP in a low
mass X-ray binary (LMXBs) is a challenging problem, both from a
theoretical and an observational point of view.  Observationally the
problem is complicated by the presence of strong timing noise, that
appears to be correlated with the X-ray flux in at least 6 AMXPs
\citep{Patruno09}. Theoretically, on the other hand, it is believed
that the NS could also be spun down during accretion if the angular
momentum transferred from the gas is not high enough, in what is known
as the ``propeller'' phase \citep{Il}. In fact recent simulations have
shown that  spin-down torques
may be present for a wide range of accretion rates \citep{rap04, Car2}. Furthermore the rapidly rotating NS may be emitting
gravitational waves (GWs) that remove angular momentum from the system
at a rate that is sufficient to halt the spin-up at a frequency of
$\approx 700$ Hz \citep{Bild98}. Finally, accretion torques
that spin-up the NS act only during relatively short outbursts,
whereas the NS is spun down by magneto-dipole torques during long
quiescence periods. Therefore the AMXP spin frequency might reach a long-term equilibrium even
if accretion torques are present during outbursts (~\citealt{Ale1}).

We present a new analysis of the torques acting
on the AMXP XTE J1814-338 (referred to as XTE J1814) and use
the results, together with those already obtained by \citet{SAX1,SAX2}
for SAX J1808.4-365 (SAX J1808), to constrain theoretical models of
the spin evolution. 

\section{Observations \& Data Analysis}

\subsection{SAX J1808.4-3658}

SAX J1808.4-3658 is a 401 Hz AMXP in a 2.01 hr binary orbit
~\citep{wij98, chak98} that has undergone six outbursts since
1996. Five outbursts have been monitored with the Rossi X-ray Timing Explorer({\it{RXTE}}).  No significant spin-up/down
episodes are detected during these outbursts, with upper limits of the
order of $|\dot{\nu}|\lesssim\,2.5\times\,10^{-14}\rm\,Hz\,s^{-1}$
\citep{SAX1,SAX2}. A long term spin-down is detected when
comparing the constant spin frequencies measured in different
outbursts.  The measured long-term spin-down is
$\dot{\nu}_{sd}\simeq\,-5.5\times\,10^{-16}\rm\,Hz\,s^{-1}$, which has
been interpreted as due to  magneto-dipole torques induced by a NS
magnetic field of $\sim\,1.5\times10^{8}$G. For this source we will use
the results reported by \citet{SAX1,SAX2} and will not perform any
new data analysis.

\subsection{XTE J1814-338}

XTE J1814 was discovered in 2003 \citep{Mark}, during an
outburst that lasted nearly 2 months. Only one outburst has been
detected so far with an extensive coverage of {\it{RXTE}} whereas a
tentative recurrence time of 19 yr has been proposed based on previous
EXOSAT observations \citep{wij03}.  Therefore the quiescence timescale
is currently unknown.  The pulsar has a spin frequency
of 314.4 Hz and orbits in 4.3 hr around a $\sim\,0.1\,M_{\odot}$
companion \citep{Mark}.  We used all available X-ray data
taken with the \textit{RXTE} Proportional Counter Array in Event mode
(time resolution $2^{-13}$s) and Good Xenon mode data (time resolution
$2^{-20}$s) in the 2.5-15 keV energy range.  We closely follow the
data reduction procedure outlined in \citet{watts08} to which we refer
for details.  A total of 28 thermonuclear X-ray bursts are detected
and removed in about 425 ks of data. Accretion-powered pulsations have
a strong overtone at twice the spin frequency \citep{wat05}.  The time
of arrivals (TOAs) of the pulsations were calculated by
cross-correlating a $\sim500$ s long folded profile with a single
sinusoid whose frequency represents the NS spin frequency. The
entire procedure is then repeated for the overtone. The effect of the
Keplerian orbital motion is removed from the time of arrivals and a
constant spin frequency is fitted to the TOAs. The X-ray flux is
calculated as the total counts in each pulse profile, normalized to
Crab units using the Crab spectral shape \citep{vanstraaten}.

\subsubsection{The Phase-Flux Correlation} 

\begin{figure}[th!]
  \begin{center}
    \rotatebox{-90}{\includegraphics[width=0.7\columnwidth]{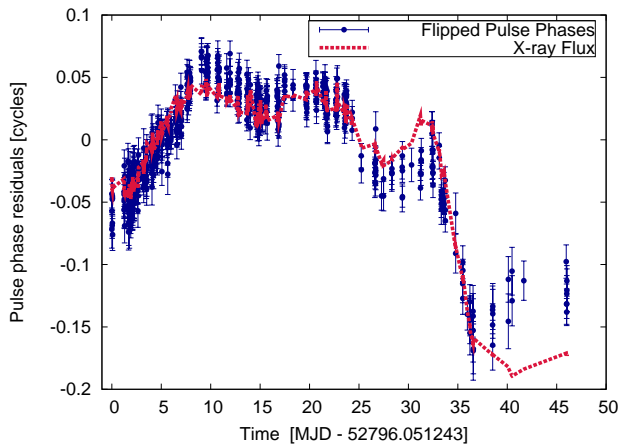}}
  \end{center}
  \caption{Phase residuals (sign reversed) of the fundamental frequency (blue dots)
    and 2.5--16 keV light curves overplotted (red line) in arbitrary units. Phase-flux correlations can be seen on
    all timescales, with small deviations at the very end of the outburst. 
    Similar results are obtained for the first overtone. \label{fig:overlap}}
\end{figure}

XTE J1814 is the AMXP that shows the strongest linear correlation
between pulse phase and X-ray flux \citep{Patruno09}. This correlation has been interpreted as possibly being generated by a moving hot spot on
the surface of the NS \citep{Patruno09, Patruno10}. Such an
interpretation is supported by recent results obtained by means of 3D
MHD simulations of AMXPs \citep{Romanova03, Bac10}. Special care must be
taken as flux induced pulse phase variations may
resemble spin frequency variations. In Fig.~\ref{fig:overlap} the effect of the phase-flux correlation is made evident by
overplotting flipped pulse phase residuals (with respect to a constant
spin frequency model) of the fundamental and the X-ray lightcurve.

It is obvious how variations of the pulse phase closely follow
variations in the X-ray flux. In standard coherent pulsar timing the spin frequency derivatives are measured by fitting a quadratic
function to the pulse phases; by doing so for XTE
J1814 the measured quadratic component will simply be a measure of
the effect that the X-ray flux has on the pulse phases.

\subsubsection{Upper limits on the Spin Frequency Derivative}

To measure the spin frequency derivative we have first removed the effect of flux variations
by using the linear phase-flux correlation (for both harmonics).
After ``cleaning'' the pulse phases, the residual rms variability is of the order of 0.02 cycles. We then fit a quadratic
function to the cleaned pulse phases and calculate the errors on the
fitted spin parameters by means of $\sim10^4$ Monte Carlo simulations
to account for the residual unmodeled variability, as described in \citet{SAX1}.

The results indicate upper limits on the spin frequency derivative of
the order of $|\dot{\nu}|\lesssim\,1.5\times\,10^{-14}\rm\,Hz\,s^{-1}$
at the $95\%$ confidence level.  The upper limits are of the same order
of magnitude as those reported by \citet{SAX1} for SAX J1808.
These results strongly contrast with the measurements obtained 
 without cleaning the pulse phases (see e.g. \citet{Papitto7}
who report $\dot{\nu}\approx -7\times\,10^{-14}\rm\,Hz\,s^{-1}$).

\section{Torque analysis}

Standard accretion theory predicts a spin-up of the form \citep{Bild97}: \be \dot{\nu}\approx
2.3\times 10^{-14}\xi^{1/2}\dot{M}_{-10}^{-5/21}B^{2/7}_8
M_{1.4}^{-3/7}R_{10}^{6/7} \mbox{Hz/s},\label{spinup}
\ee where $\dot{M}_{-10}$ is the accretion rate in units of $10^{-10}
M_\odot/\mbox{yr}$, $B_{8}$ the magnetic field in units of $10^8$ G,
$M_{1.4}$ the mass of the NS in units of $1.4 M_{\odot}$, $R_{10}$ its
radius in units of $10$ km and $\xi$ parametrises the
uncertainties in evaluating the torque at the edge of the accretion
disc and is thought to be in the range $\xi\approx 0.3-1$
\citep{Psaltis}.  
The average accretion rate for the outburst is $\simeq\,5\times
10^{-10}\rm\,M_{\odot}\,yr^{-1}$ and
$2\times10^{-10}\rm\,M_{\odot}\,yr^{-1}$ for SAX J1808 and XTE J1814,
respectively, when considering the bolometric flux (using the data reported by \citealt{Heinke, wij03}). The expression in
(\ref{spinup}) would thus give $\dot{\nu}\simeq\,9\times10^{-14}\rm\,Hz\,s^{-1}$ for SAX J1808 and
$\dot{\nu}\simeq 4\,10^{-14}\rm\,Hz\,s^{-1}$ for XTE J1814, both larger than the observational upper limits.

Our results suggest that during the outburst SAX J1808 and XTE J1814 spin up/down much less than can be measured and that the accretion torque is thus much weaker than the estimate in (\ref{spinup}), or that there is an additional spin-down torque acting on the star, due for example to GW emission.

\subsection{Gravitational wave torques}

Gravitational wave emission was first suggested as the cause for the
cutoff in the spin distribution of the LMXBs more than thirty years
ago \citep{PP}. The main emission mechanisms that could be at work in these
systems are crustal ``mountains'' \citep{Bild98,UCB}, magnetic deformations
\citep{Cutler, Melatos} or unstable modes \citep{Nils98}. All these
processes can produce a substantial quadrupole $Q_{22}$ and thus a spin-down torque due to GW emission.

\subsubsection{Crustal mountains}

The crust of a NS is a thin layer ($\approx$ 1 km thick) of matter in
a crystalline phase, that can support shearing and the presence
of a small asymmetry, or ``mountain''. The crust consists of
several layers of different nuclear composition and as accreted matter
gets pushed further into the star it undergoes a series of nuclear
reactions, including electron captures, neutron emission and
pycnonuclear reactions \citep{Sato, Haensel1}. The heat deposited in
the whole crust due to these reactions is $Q_H\approx 1.5$ MeV per
accreted baryon, but most of the heat is released by reactions
occurring close to neutron drip \citep{Haensel1}. These reactions will
heat the region by an amount \citep{UR}:
\be
\delta T\approx 10^3 {C}_{k}^{-1}{p}_{30}^{-1}{Q_n}{\Delta M}_{22}\;\mbox{K},
\label{delta1}
\ee 
where $C_{k}$ is the heat capacity in units of $k_B$ (the Boltzmann constant) per
baryon, $p_d$ is the pressure at which the reactions occur (in units of $10^{30}$ erg/cm$^3$) and $Q_n$
the deposited heat per accreted baryon in MeV. $\Delta M$ is the accreted
mass in units of $10^{22}$ g, which for the systems we consider
is approximately the amount of matter accreted after a few days of
outburst.  If the energy deposition is (partly) asymmetric this would
perturb the equilibrium stellar structure and give rise to a mass
quadrupole\citep{UCB}: \be Q_{22}\approx 1.3 \times
10^{35} R_6^4\left(\frac{\delta
    T_q}{10^5\mbox{K}}\right)\left(\frac{Q}{30\mbox{MeV}}\right)^3
\mbox{g cm$^2$}\label{delta2}, \ee where $\delta T_q$ is the asymmetric
(quadrupolar) part of the temperature increase and $Q$ is the
threshold energy for the reaction. The quadrupole
required for spin equilibrium during an outburst is $Q\approx 10^{37}$ g cm$^2$ for both systems \citep{UCB}. It is clear from equations (\ref{delta1},\ref{delta2}) that even under the most optimistic assumptions it is
very unlikely to build a ``mountain'' large enough to balance spin-up during accretion. 

\subsubsection{Magnetic mountains}

It is well known that a magnetic star will not be
spherical and, if the rotation axis and the magnetic axis
are not aligned, one could have a ``magnetic mountain'' leading to GW
emission. However such deformations are unlikely to be large enough to balance the accretion torques in weakly
magnetised systems such as the LMXBs \citep{magnetic}. Large internal fields could also cause a deformation \citep{Cutler}, but this would persist in quiescence, leading to a rapid spin-down, of the order of the spin-up in (\ref{spinup}),
which is not observed in SAX J1808 \citep{SAX2}.

Another possibility is that the magnetic field lines are stretched by
the accreted material itself as it spreads on the star, giving rise to
a large magnetically confined mountain. The results of \citet{Melatos}
suggest that the quadrupole built this way could balance the
accretion torque only if the surface field is significantly stronger than
the external dipole component. Furthermore such a mountain would persist on an ohmic dissipation timescale $\tau_{_{Ohm}}\approx 10^2$ yrs \citep{Melatos}
and thus should also give rise to a strong additional spin-down in
quiescence.

\subsubsection{Unstable modes}

A more promising scenario is that of an oscillation mode of the NS
being driven unstable by GW emission, the main candidate for this mechanism
being the $l=m=2$ r-mode\citep{Nils98}. An r-mode
is a toroidal mode of oscillation for which the restoring force is the
Coriolis force. It can be driven unstable by the
emission of GWs, as long as viscosity does not damp it on a
faster timescale. This will only happen in a narrow window in
frequency and temperature which depends on the microphysical
details of the damping mechanisms (for a review see \citet{review}).

Of particular interest in this context is the situation in which the
frequency at which the mode is driven unstable (the 'critical'
frequency) increases with temperature in the range we are considering
($T\approx 10^7-10^8$ K). This could lead to the system being in
equilibrium at the critical frequency and emitting GWs at a level that
will balance the accretion torque \citep{NilsST}. This would
be the scenario if the core contains hyperons \citep{Nayyar, hyperon}
or deconfined quarks \citep{NilsST}.  If this is the case the
dimensionless amplitude of the mode $\alpha$ (as defined by
\citealt{Owen}) will be approximately constant in time,
i.e. $\dot\alpha\approx 0$, and if we assume that GW emission is
balancing the accretion torque one has:
\be\alpha\approx 1.9\times 10^{-6} \dot{M}_{-10}^{3/7}\nu_{400}^{-7/2}\xi^{1/4} B_8^{1/7},\ee
where $\nu_{400}=\nu/400\mbox{ Hz}$ and we have assumed a 1.4 $M_\odot$ mass and 10 km radius, which we take as standard values from now on.

After the outburst, when accretion ceases, the system will cool and
spin-down following the critical frequency curve. Given that
$\dot{\alpha}\approx 0$ the mode amplitude would be approximately
constant resulting in a spin-down rate $\dot{\nu}\approx -7\times
10^{-14}$ Hz s$^{-1}$ for both systems. This is considerably higher
than the observed quiescent spin-down rate for SAX J1808 of
$\dot{\nu}\approx -5.5\times 10^{-16}$ Hz s$^{-1}$.

Furthermore the shear from the mode will heat the star at a rate \citep{review}:
\be
\dot{E}\approx 1.8\times 10^{44}\alpha^2 T_8^{-5/3}\nu_{400}^2\mbox{erg s$^{-1}$},\label{heat}
\ee
where $T_8=(T/10^8 K)$. Let us first of all assume that the core cools due to the
modified URCA neutrino emission process, which leads to an energy loss
$\dot{E}\approx 8.4\times 10^{31}T_8^8$ erg/s
\citep{Page}. Balancing this energy loss with the viscous r-mode
heating in (\ref{heat}) leads to an equilibrium temperature of
$T\approx 1.4 \times 10^{8}$ K for SAX J1808 and $T\approx 1.5 \times
10^{8}$ K for XTE J1814.  Let us now estimate the temperature in the
presence of the strongest possible neutrino emission processes. We therefore assume that nucleon direct URCA processes are
possible, which may be the case in SAX J1808 \citep{Yak}, and that the core is not superfluid, with $\dot{E}_\nu\approx
4\times 10^{39} T_8^6$ erg/s \citep{Page} (similar processes would be at work if hyperons or deconfined quarks are present). Note that in the presence of
superfluidity these processes  would be strongly
quenched (for a recent analysis of r-mode heating see \cite{Wynn}). The temperatures we obtain for the stars are $T\approx 1.5\times 10^7$ K for SAX J1808 and
$T\approx 1.6\times 10^7$ K for XTE J1814.

How do these temperatures compare with observational constraints?
Both systems have upper limits on their
temperature from quiescent observations \citep{Heinke}. One can map
the surface temperature to the temperature at the top of the crust via
the relation given by \citet{Pot} for an accreted crust: \be
\left(\frac{T_{eff}}{10^6
    \mbox{K}}\right)^4=\left(\frac{g}{10^{14}\mbox{cm
      s$^{-2}$}}\right)\left(18.1\frac{T}{10^9 \mbox{K}}\right)^{2.42},
\ee where $g$ is the gravitational acceleration at the surface. By
assuming that the star is isothermal, which is roughly correct if the
thermal conductivity is high, as observations of cooling X-ray
transients suggest \citep{Brown09}, one can then obtain a core
temperature of $T<8.6 \times 10^6$ K for SAX J1808 and $T< 3.4 \times
10^7$ for XTE J1814.

It would thus appear that the presence of an unstable r-mode is not
consistent with the observed quiescent spin-down of
SAX J1808, while it could be marginally consistent with the
temperature of both sources only in the presence of direct URCA processes in the core.

\subsection{Accretion torques}

The interaction between the accretion disc and the magnetosphere is
the natural candidate to explain not only the behaviour of
our two systems, but also the cutoff in the
spin-distribution of the millisecond pulsars. This mechanism was
considered in detail by \citet{WZ}, whose study seemed to indicated an
unexpected link between the accretion rate and the magnetic field
strength. This led to the suggestion that GW torques may be
active, as in this case the dependence of the field strength on the
accretion rate is weaker.

Recent work has, however, shown that a more detailed modelling of magnetic torques in the disc can alleviate this problem \citep{NilsSpin}. It could, therefore, be the case that
both SAX J1808 and XTE J1814 are close to the equilibrium spin rate
set by such a process and that the torques are thus much weaker than
the estimate in (\ref{spinup}).  In particular we will use the simple
model of \citet{NilsSpin} to show that this is a more promising
explanation than GW emission.

\begin{figure}[th!]
  \begin{center}
    \rotatebox{-90}{\includegraphics[width=0.7\columnwidth]{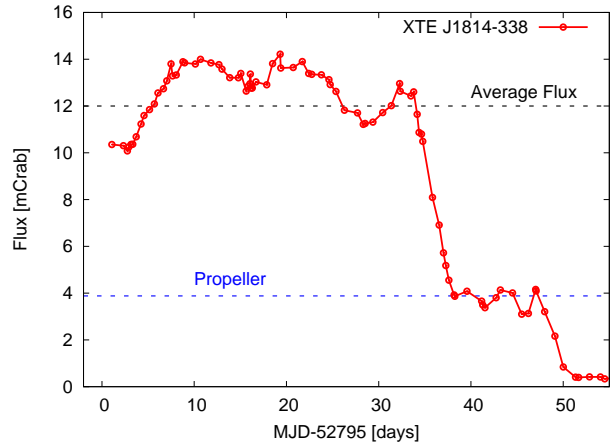}}
  \end{center}
  \caption{X-ray lightcurve [2.5--16 keV] of XTE J1814. The dashed blue line indicates the possible onset of the propeller ($r_m=r_c$) while the dashed black line is the average flux, which is close to the equilibrium condition $r_m\approx 0.8\,r_{c}$.
\label{fig2}}
\end{figure}

We assume that a propeller phase starts
at the reflaring stage for SAX J1808 (see a discussion of this
possibility in \citealt{pat09}). In XTE J1814 we assume that the onset
of the propeller is at the point where the pulse phases
deviate from a linear correlation (see Fig~\ref{fig:overlap} and
~\ref{fig2}).

We calculate the average X-ray flux during an outburst and the flux at
the onset of the propeller for XTE J1814 (see Fig~\ref{fig2}) and SAX
J1808 and assume that the X-ray flux is a good tracer of the mass
accretion rate. We assume that at the onset of the propeller the magnetospheric radius \citep{Bild97}:
\be
r_m=35 \xi \dot{M}_{-10}^{-2/7} M^{-1/7}_{1.4}
R_{10}^{12/7}B_8^{4/7} \mbox{km} 
\ee
is approximately equal to the
co-rotation radius 
\be
r_c=31 M_{1.4}^{1/3}\nu_{400}^{-2/3}
\mbox{km}.
\ee
The results of \citep{NilsSpin} indicate that the torque will vanish when the ratio between the propeller flux ($f_{prop}$) and the
average flux ($f_{avg}$) is such that $(f_{prop}/f_{avg})^{2/7}\approx 0.8$. We find that ($(f_{prop}/f_{avg})^{2/7}\approx 0.75$ for XTE
J1814 and 0.75-0.84 for SAX J1808 in the 2002, 2005 and 2008
outbursts\footnote{Only these outbursts had good enough observational coverage to allow the calculation.}.  These values are very close to the theoretical value
$0.8$, which is encouraging given the large uncertainties in both the
theoretical calculations and the observations.

This does not mean that the accretion torques vanish during
an outburst, but only that the spin variations can be smaller than
those predicted by standard accretion theory. 

\section{Conclusions}

We have presented a new analysis of the torques acting
on XTE J1814 during the 2003 outburst and have interpreted the observed
long term spin-down of SAX J1808 over a 10 yr baseline.  We have found
that there are no significant spin frequency derivatives that can be
measured during the 2003 outburst of XTE J1814, with upper limits of
$|\dot{\nu}|\lesssim1.5\times\,10^{-14}\rm\,Hz\,s^{-1}$, of the same
order of magnitude as the values determined for SAX J1808:
$|\dot{\nu}|\lesssim2.5\times\,10^{-14}\rm\,Hz\,s^{-1}$~\citep{SAX1,SAX2}.
This implies that accretion torques during an outburst are smaller
than expected from standard accretion theory for both AMXPs.

We have analysed the two main mechanisms
that could explain such a feature, namely the presence of GW torques
and the possibility that the system may be close to spin equilibrium,
as set by the disc/magnetosphere interaction. For SAX J1808 it is
reasonably safe to say that GW torques can be excluded.  For XTE J1814 a definite conclusion is harder to draw,
as only one outburst has been observed and thus we do not have a
measurement of the spin-down in quiescence. GW emission may still be marginally consistent with observations.

A more promising possibility is that both systems are close to spin
equilibrium, as set by the disc/magnetosphere interaction. We have
shown that the simple model of \citet{NilsSpin} predicts spin
equilibrium close to the average accretion rate for the outbursts,
thus leading to a reduced torque. Further work is necessary to analyse
the problem with the use of refined disc models (e.g.
\citealt{Car2}), a task which is beyond the scope of this
letter.

Note that the case of bright sources
such as Sco X-1 could be quite different, as high accretion rates would prevent the system from reaching spin equilibrium and GWs may play a significant role.



\begin{thebibliography}{38}

\bibitem[\protect\citeauthoryear{Andersson}{Andersson}{1998}]{Nils98} Andersson N., 1998, Ap.J. 502, 708
\bibitem[\protect\citeauthoryear{Andersson, Kokkotas \& Ferrari}{Andersson, Kokkotas \& Ferrari}{2001}]{review} Andersson N., Kokkotas K.D., Ferrari V., 2001, Int.Journ.Mod.Phys.D 10, 381
\bibitem[\protect\citeauthoryear{Andersson, Jones \& Kokkotas}{Andersson, Jones \& Kokkotas}{2002}]{NilsST} Andersson N., Jones D.I., Kokkotas K.D., 2002, MNRAS 337, 1224
\bibitem[{{Andersson} {et~al.}(2005){Andersson}, {Glampedakis}, {Haskell}, \&
  {Watts}}]{NilsSpin}{Andersson}, N., {Glampedakis}, K., {Haskell}, B., \& {Watts}, A.~L. 2005,
  \mnras, 361, 1153
\bibitem[Bachetti et al.(2010)]{Bac10} Bachetti, M., Romanova, M.~M., 
Kulkarni, A., Burderi, L., di Salvo, T., \ 2010, \mnras, 403, 1193
\bibitem[{{Bildsten} {et~al.}(1997){Bildsten}, {Chakrabarty}, {Chiu} et al.}]{Bild97}
{Bildsten}, L., Chakrabarty, D., Chiu {et~al.} 1997, \apjs, 113, 367
\bibitem[\protect\citeauthoryear{Bildsten}{1998}]{Bild98} Bildsten L., ApJ.Lett. 501, L89

\bibitem[\protect\citeauthoryear{Brown \& Ushomirsky}{2000}]{BU98} Brown E.F., Ushomirsky G., 2000, \apj 536, 915
\bibitem[\protect\citeauthoryear{Brown \& Cumming}{2009}]{Brown09} Brown E., Cumming A., 2009, \apj 698, 1020

\bibitem[\protect\citeauthoryear{Chakrabarty, Morgan, Muno, Galloway, Wijnands, van der Klis \& Markwardt}{Chakrabarty et al.}{2003}]{Chak1} Chakrabarty D., Morgan E.H., Muno M.P. et al., 2003, Nature 424, 42
\bibitem[\protect\citeauthoryear{Cutler}{Cutler}{2002}]{Cutler} Cutler C., 2002, Phys.Rev.D 66, 084025
\bibitem[\protect\citeauthoryear{D'Angelo \& Spruit}{2011}]{Car2} D'Angelo C.~R. \& Spruit H.~C.,2011, preprint arXiv:1102.3697


\bibitem[Chakrabarty \& Morgan(1998)]{chak98} Chakrabarty, D., \& Morgan, E.~H.\ 1998, \nat, 394, 346 

\bibitem[\protect\citeauthoryear{Patruno, Wijnands \& van der Klis}{2009}]{Patruno09} Patruno A., Wijnands R., van der Klis, M., 2009, Ap.J. 698, L60
\bibitem[\protect\citeauthoryear{Patruno}{2010}]{Ale1} Patruno A., 2010, Ap.J. 722, 909


\bibitem[\protect\citeauthoryear{Haensel \& Zdunik}{Haensel \& Zdunik }{1990}]{Haensel1} Haensel P., Zdunik J.L., 1990, A\&A 227, 431
\bibitem[\protect\citeauthoryear{Hartman et al.}{2008}]{SAX1} Hartman J.M., Patruno A., Chakrabarty D. et al., ApJ 675, 1468
\bibitem[\protect\citeauthoryear{Hartman et al.}{2009}]{SAX2} Hartman J.M., Patruno A., Chakrabarty D. et al., ApJ 702, 1673
\bibitem[\protect\citeauthoryear{Haskell, Samuelsson, Glampedakis \& Andersson}{Haskell et al.}{2008}]{magnetic} Haskell B., Samuelsson L., Glampedakis K.,Andersson N., 2008, \mnras, 385, 531
\bibitem[\protect\citeauthoryear{Haskell \& Andersson}{Haskell \& Andersson}{2010}]{hyperon} Haskell B., Andersson N., 2010, MNRAS 408, 1897

\bibitem[\protect\citeauthoryear{Heinke, Jonker, Wijnands, Deloye \& Taam}{Heinke et al.}{2009}]{Heinke} Heinke C.O, Jonker P.G., Wijnands R., Deloye C.J., Taam R.E., 2009, \apj 691, 1035
\bibitem[\protect\citeauthoryear{Ho, Andersson \& Haskell}{Ho et al.}{2011}]{Wynn} Ho W.C.G., Andersson N., Haskell B., 2011, in preparation
\bibitem[\protect\citeauthoryear{Illarionov \& Sunyaev}{1975}]{Il} Illarionov, A.~F., Sunyaev R.~A.,1975, A\&A, 39, 185
\bibitem[\protect\citeauthoryear{Markwardt \& Swank}{2003}]{Mark} Markwardt C., Swank, 2003, IAU circ. 8144, 1 

\bibitem[\protect\citeauthoryear{Melatos \& Payne}{2005}]{Melatos} Melatos A., Payne D.J.B., 2005, \apj 623, 1044
\bibitem[\protect\citeauthoryear{Nayyar \& Owen}{Nayyar \& Owen}{2006}]{Nayyar} Nayyar M., Owen B.J., 2006, Phys.Rev.D 73, 084001
\bibitem[\protect\citeauthoryear{Owen, Lindblom, Cutler, Schutz, Vecchio \& Andersson}{Owen et al.}{1998}]{Owen} Owen B.J., Lindblom L., Cutler C., Schutz B.F. et al., 1998, Phys.Rev.D. 58, 084020
\bibitem[\protect\citeauthoryear{Page, Geppert \& Weber}{2006}]{Page} Page D.,Geppert U., Weber F., 2006, Nucl.Phys.A 777, 497
\bibitem[\protect\citeauthoryear{Papaloizou \& Pringle}{1978}]{PP} Papaloizou J., Pringle J.E., 1978, MNRAS 184, 501
\bibitem[\protect\citeauthoryear{Papitto et al.}{2007}]{Papitto7} Papitto A., di Salvo T., Burderi L., Menna M.T., Lavagetto G., Riggio A., 2007, \mnras 375, 971 

\bibitem[Patruno et al.(2009)]{pat09} Patruno, A., Watts, A., 
Klein Wolt, M., Wijnands, R., \& van der Klis, M.\ 2009, \apj, 707, 1296 
\bibitem[Patruno et al.(2010)]{Patruno10} Patruno, A., Hartman, J.~M., Wijnands, R., Chakrabarty, D., 
\& van der Klis, M.\ 2010, \apj, 717, 1253 


\bibitem[{{Psaltis} \& {Chakrabarty}(1999)}]{Psaltis}
{Psaltis}, D., \& {Chakrabarty}, D. 1999, \apj, 521, 332

\bibitem[\protect\citeauthoryear{Potekhin, Chabrier \& Yakovlev}{Potekhin, Chabrier \& Yakovlev}{1997}]{Pot} Potekhin A.Y., Chabrier G., Yakovlev D.G., 1997, A\&A 323, 415

\bibitem[Rappaport et al.(2004)]{rap04} Rappaport, S.~A., 
Fregeau, J.~M., \& Spruit, H.\ 2004, \apj, 606, 436 
\bibitem[Romanova et al.(2003)]{Romanova03} Romanova, M.~M., 
Ustyugova, G.~V., Koldoba, A.~V., Wick, J.~V., \& Lovelace, R.~V.~E.\ 2003, \apj, 595, 1009 
\bibitem[\protect\citeauthoryear{Sato}{1979}]{Sato} Sato K., 1979, Prog.Theor.Physics 62, 957


\bibitem[\protect\citeauthoryear{Ushomirsky, Cutler \& Bildsten}{Ushomirsky, Cutler \& Bildsten}{2000}]{UCB} Ushomirsky G., Cutler C., Bildsten L., 2000, MNRAS 319, 902
\bibitem[\protect\citeauthoryear{Ushomirsky \& Rutledge}{2001}]{UR} Ushomirsky G., Rutledge R.E., 2001, MNRAS 325, 1157
\bibitem[van Straaten et al.(2003)]{vanstraaten} van Straaten, S., 
van der Klis, M., \& M{\'e}ndez, M.\ 2003, \apj, 596, 1155 
\bibitem[Watts et al.(2005)]{wat05} Watts, A.~L., Strohmayer, 
T.~E., \& Markwardt, C.~B.\ 2005, \apj, 634, 547 
\bibitem[Watts et al.(2008)]{watts08} Watts, A.~L., Patruno, 
A., \& van der Klis, M.\ 2008, \apjl, 688, L37 

\bibitem[\protect\citeauthoryear{White \& Zhang}{1997}]{WZ} White N., Zhang W., 1997, \apj 490, L87
\bibitem[Wijnands \& van der Klis(1998)]{wij98} Wijnands, R., \& van der Klis, M.\ 1998, \nat, 394, 344 
\bibitem[Wijnands \& Reynolds(2003)]{wij03} Wijnands, R., \& Reynolds, A.\ 2003, The Astronomer's Telegram, 166, 1
\bibitem[Yakovlev, Levenfish \& Haensel (2003)]{Yak} Yakovlev, D.G., Levenfish, K.P., Haensel P., 2003, A\& A, 407, 265 

  \end{thebibliography}
\end{document}